\documentclass{ws-rv9x6}
\usepackage{subfigure}   
\usepackage{ws-rv-thm}   
\usepackage{ws-rv-van}   
\makeindex
\usepackage[colorlinks=true, urlcolor=blue, anchorcolor=blue, citecolor=blue,filecolor=blue,linkcolor=blue,menucolor=blue,pagecolor=blue]{hyperref}

\usepackage[normalem]{ulem}
\usepackage{amsmath,amssymb,graphicx,cases}
\usepackage{epsfig}
\usepackage{textcomp}
\usepackage{epstopdf}
\usepackage{float}
\renewcommand{\vec}[1]{\boldsymbol{#1}}

\begin{document}

\chapter*{Survival, absorption and escape of interacting diffusing particles}\label{ra_ch1}

\author{Tal Agranov  and Baruch Meerson}

\address{Racah Institute of Physics, Hebrew University of
	Jerusalem, Jerusalem 91904, Israel}

\begin{abstract}
At finite concentrations of reacting molecules,  kinetics of diffusion-controlled reactions is affected by intra-reactant interactions. As a result, multi-particle reaction statistics cannot be deduced from  single-particle results. Here we briefly review a recent progress in overcoming this fundamental difficulty. We show that the fluctuating hydrodynamics and macroscopic fluctuation theory provide a simple, general and versatile framework for studying a whole class of problems  of survival, absorption and escape of interacting diffusing particles.
\end{abstract}
\markright{Survival, absorption and escape} 
\body

\tableofcontents

\section{Introduction}\label{1}
Kinetics of many diffusion-controlled reactions is affected by intra-reactant interactions. This happens when the density of the reacting molecules is not too small.  Although the importance of interactions may have been recognized for a long time, there has been very little progress in their account in theory. Here we will briefly review, and slightly generalize, one promising approach toward solving this long-standing problem \cite{MVK,surv,fullabsorb,multi,narrow}. We will consider several prototypical gas settings.  The first group of settings -- interior settings -- deals with interacting diffusing molecules inside a domain (think about a living cell).  The second group -- exterior settings -- deals with molecules surrounding a domain. In both cases the domain boundary, or part of it, absorbs the molecules upon impact, signaling that a reaction occurred.

\begin{figure} [ht]\label{hure}
 	\includegraphics[width=0.49\textwidth,clip=]{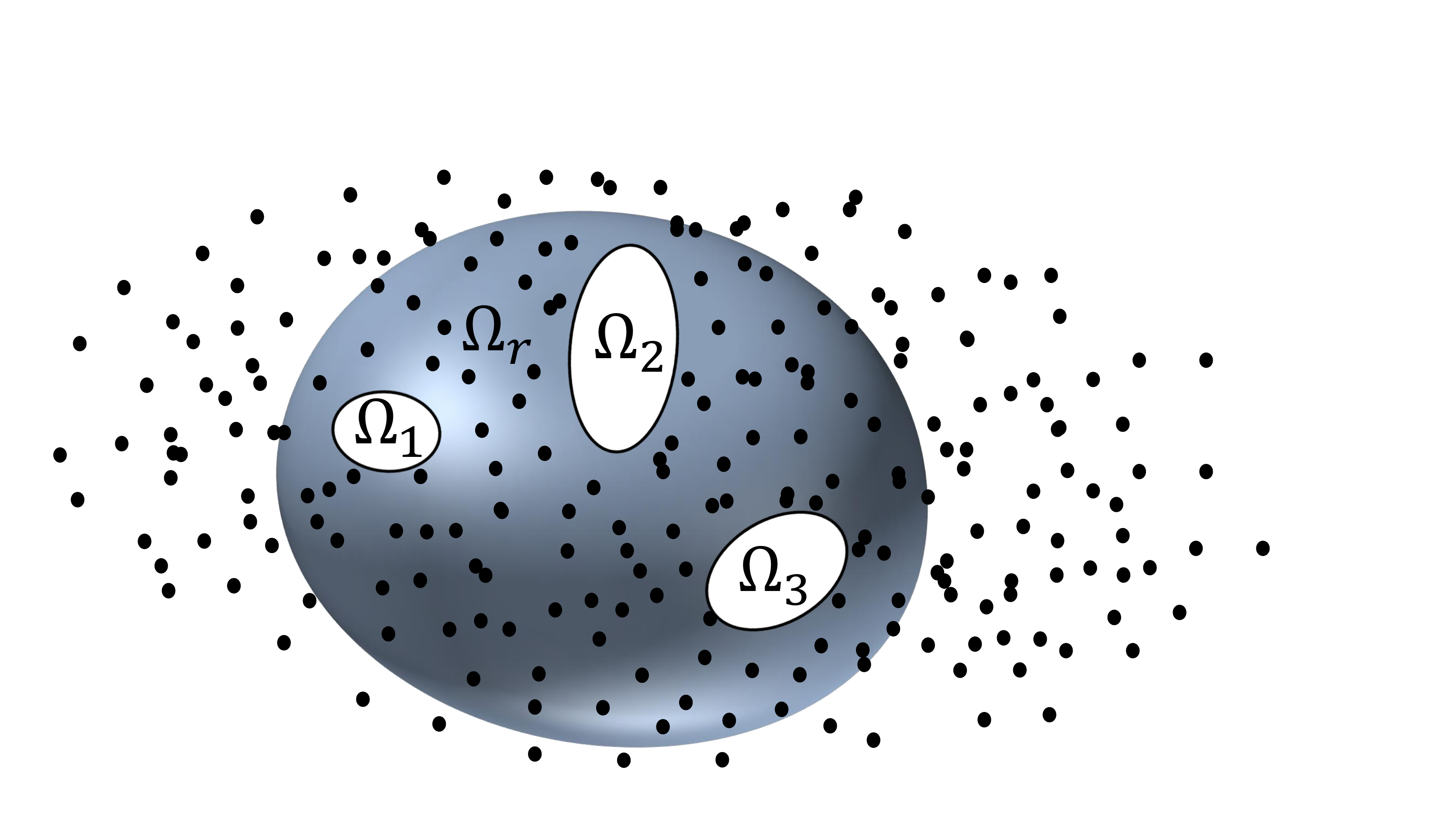}
  	\includegraphics[width=0.49\textwidth,clip=]{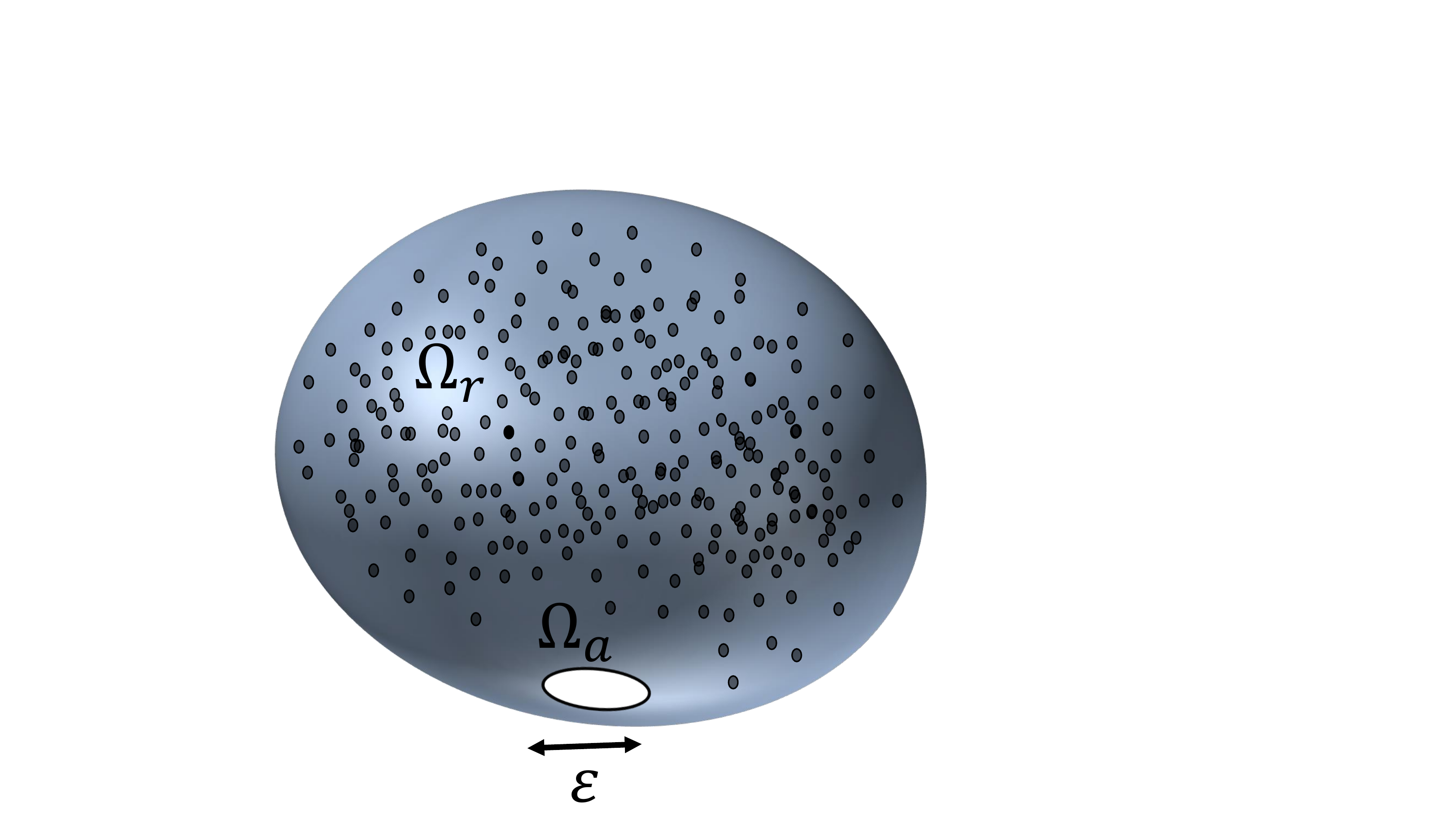}
 	\caption{Left panel: An exterior problem with multiple absorbing patches shown in white. The rest of the boundary is reflecting. Right panel: Narrow escape of multiple particles through a small hole in the boundary, the rest of the boundary being reflecting. }
 \end{figure}

The interior settings give simplified descriptions of inter-cellular transport in the living cell, where molecules search for a correct location within a cell  membrane. The efficacy of the inter-cellular transport is determined by the absorption rate of the molecules \cite{bress}. The interior settings include the \emph{narrow escape problem}, see the right panel of Fig.\ref{hure}, which is well studied in the case of non-interacting diffusing molecules trying to escape from a closed domain via a small hole in its boundary \cite{keller1,bress,beni,bookz,rev,russians}.

The exterior problems are different but closely related. The case when the boundary of the domain is fully absorbing is known as the target search, or target survival problem \cite{rice,osh,red}. This describes the situation where the molecules of one reactant -- a minority -- can be viewed as big and immobile, whereas the molecules of another reactant  -- a majority -- are small and mobile. A different scenario happens when molecules get absorbed only through some absorbing patches -- receptors -- distributed on the otherwise reflecting domain boundary \cite{berg,bergbook}, see the left panel of Fig. \ref{hure}.

If the diffusing molecules are treated as noninteracting random walkers, the calculation of the effective reaction rates and its fluctuation  statistics boils down to calculating a single-particle probability.  Interactions invalidate the single-particle picture and make the problem very difficult.  Fortunately, a new simplification emerges if there are sufficiently many interacting diffusing particles in the relevant region of space. In this case one can use the fluctuating hydrodynamics, which goes back to Landau and Lifshitz \cite{LL}, and a large deviation theory for it. As a convenient and well controlled first-principle model, Refs. \cite{MVK,surv,fullabsorb,multi,narrow} adopted
diffusive lattice gases, where the fluctuating hydrodynamics is well established \cite{Spohn}. The corresponding large-deviation theory has recently become available under the name of macroscopic fluctuation theory (MFT) \cite{MFTreview}.

\section{Fluctuating Hydrodynamics and the MFT}
Fluctuating hydrodynamics is a coarse-grained description of a gas of particles in terms of the particle number density $\rho(\mathbf{x},t)$ \cite{Spohn,KL}. The average particle density of a lattice gas obeys a diffusion equation $\partial_t \rho = \nabla \cdot \left[D(\rho) \nabla \rho\right]$, whereas macroscopic fluctuations are described by the conservative Langevin equation
\begin{equation}
\partial_t \rho = -\nabla \cdot \mathbf{J}\,,\quad\mathbf{J}=-D(\rho) \nabla \rho-\sqrt{\sigma(\rho)}\boldsymbol{\eta}(\mathbf{x},t),\label{lang}
\end{equation}	
where $\boldsymbol{\eta}(\mathbf{x},t)$ is a zero-mean Gaussian noise, delta-correlated in space and in time \cite{Spohn,KL,Liggett}.  The diffusivity $D(\rho)\geq 0$ and the mobility $\sigma(\rho)\geq 0$ are to be obtained, for each lattice gas, from the microscopic model. The simplest case, whose coarse-grained behavior coincides with that of non-interacting Brownian particles is the gas of Random Walkers (RWs)\cite{Paulbook}, where one has $D(\rho)=D_0$ and $\sigma(\rho)=2D_0\rho$ \cite{Spohn}. A model with interactions, which we will focus on, is the symmetric simple exclusion process (SSEP) \cite{Spohn} which accounts, in a simple way, for excluded volume interactions. The SSEP's average behavior coincides with that of the RW's, as they share the same density-independent diffusivity $D_0$. Their fluctuations, however, are different as the SSEP's mobility  $\sigma(\rho)=2D_0\rho(1-\rho a^3)$ is a non-linear function of $\rho$ \cite{Spohn,KL}. Here $a$ is the lattice constant which we set to unity, so that  $0\leq\rho\leq1$. Like many other lattice gases, the SSEP behaves in its dilute limit as non-interacting RWs.

To develop a large-deviation theory for Eq. (\ref{lang}), one starts from a path integral for the probability of observing a joint density and flux histories $\rho(\mathbf{x},t), \mathbf{J}(\mathbf{x},t)$, constrained by the continuity equation (\ref{lang}):
\begin{eqnarray}\nonumber
&&\!\!\!\!\!\!\mathcal P =\int\mathcal{D}\rho\mathcal{D}\mathbf{J}
\prod_{\mathbf{x},t}\delta(\partial_{t}{\rho}+\nabla \cdot \mathbf{J})\,
\exp\left(-\mathcal{S}\right),\label{path0}\\
&&\!\!\!\!\!\!\mathcal{S}\left[\rho(\mathbf{x},t),\mathbf{J}(\mathbf{x},t)\right]=\int_0^Tdt\int d^3\mathbf{x}\frac{\left[\mathbf{J}+D(\rho)\nabla \rho\right]^2}{2\sigma(\rho)}.\label{path1}
\end{eqnarray}
In the interior and exterior settings, presented above,  we condition the process on a specified (zero or non-zero) particle absorption current by a given time $T$. Therefore, we need to evaluate the path integral over only those density and flux histories which led to the specified current. Assuming that all characteristic length scales involve large numbers of particles, the dominant contribution for $\mathcal P$ comes from the \emph{optimal fluctuation}: the most probable history $\rho(\mathbf{x},t), \,\mathbf{J}(\mathbf{x},t)$ \cite{MSR}.
The ensuing minimization procedure yields the Euler-Lagrange equation which can be cast into a Hamiltonian form, known as the MFT equations \cite{MFTreview}. The minimization procedure also generates problem-specific boundary conditions. Evaluating the minimum action $S$ over the solutions to the minimization problem yields the desired probability ${\mathcal P}$ up to a pre-exponential factor,
\begin{eqnarray}
\label{actionmain}
-\ln {\mathcal P}
\simeq S \equiv\min_{\rho,\mathbf{J}}\mathcal{ S}\left[\rho(\mathbf{x},t),\mathbf{J}(\mathbf{x},t)\right].
\end{eqnarray}
In general, the minimization problem is not solvable analytically. Considerable simplifications arise in the limits of very long and very short times compared to a characteristic diffusion time of the problem, see below. We will first address the long-time limit, where the optimal gas density and flux become \emph{stationary}, and devote the last chapter \ref{nonstat} to the non-stationary regime relevant for short times. Following Refs. \cite{MVK,surv,fullabsorb,multi,narrow}, we will consider two types of initial conditions. The first is a random (or annealed) initial condition, where particles are randomly distributed in space with an average density $\rho_0$. It describes the situation where the gas has enough time to equilibrate before the process starts (for example, before the receptor becomes available). The other is deterministic (or quenched) initial condition with a uniform density $\rho_0$. When considering long times the details of the initial condition become irrelevant. In contrast, the short-time statistics strongly depends on the initial condition.

\section{The exterior problem}
Suppose a gas of diffusing particles fills the whole space outside of a simply connected $3d$ domain of a linear size $L$.
The domain boundary $\Omega$ is composed of a reflecting part $\Omega_r$ and a complementary absorbing part $\Omega_a$. Whenever a particle hits $\Omega_a$, it is immediately absorbed, which sets
\begin{eqnarray}\label{blang}
&&\rho(\mathbf{x}\in{\Omega_a},t)=0.
\end{eqnarray}
Whenever a particle hits $\Omega_r$, it is reflected, which sets a zero-flux boundary condition
\begin{eqnarray}\label{blang2}
&&\mathbf{J}(\mathbf{x}\in{\Omega_r},t)\cdot\hat{n}=0,
\end{eqnarray}
where $\hat{n}$ denotes a local unit vector normal to the domain boundary and directed into the domain.
For a fully absorbing domain there is no reflecting part. The simpler latter setting is known as ``the target search problem" \cite{rice,osh,red}; it captures the essence of many diffusion-controlled chemical reactions.
A more involved setting is a domain whose boundary has several disjoint absorbing patches (receptors) \cite{berg,bergbook}.
For both random and deterministic initial conditions the boundary condition at infinity is
 \begin{eqnarray}\label{blang3}
 \rho(|\mathbf{x}|\rightarrow\infty,t)=\rho_0.
 \end{eqnarray}
The quantity of interest is the probability $\mathcal{P}(n,T,\rho_0)$ that $N$ gas particles were absorbed during the time interval $0<t<T$, where $n\equiv N/T$ is the absorption current. For multiple absorbing patches one is interested in the corresponding multivariate probability.

At times $T$ much longer than the diffusion time $L^2/D(\rho_0)$ the system reaches a non-equilibrium steady state, where the average gas density $\bar{\rho}(\mathbf{x})$ is independent of time. In its turn, the average number of absorbed particles, $\bar{N}$, is proportional to time, so that the
average absorption current $\bar{n}=\bar{N}/T$ is independent of time. Similarly, for a whole class of lattice gases the \emph{optimal} density and flux, conditioned on a specified current $n\neq\bar{n}$, also become stationary \cite{layers}.
As a result, ${\mathcal P}(n,T,\rho_0)$  exponentially decays with time $T$. A similar situation occurs in the context of stationary fluctuations of current in diffusive lattice gases, driven by density reservoirs at the boundaries. There the stationarity of the optimal gas density and flux is known under the name of the ``additivity principle"  \cite{bd}, and we use this term here as well.

\subsection{Target survival}\label{exsurv}
The authors of\cite{MVK} considered a fully absorbing domain ($\Omega_a=\Omega$) and studied the probability $\mathcal{P}(n=0,T,\rho_0)$ that not a single particle  gets absorbed by time $T$. This probability is often called the survival probability; it is a key quantity in determining the distribution of absorption times of the first particle. The latter is given by $\mathcal{P}_{\text{first}}(T,\rho_0)=-\partial_T\mathcal{P}(n=0,T,\rho_0)$. As a result, the mean absorption time of the first particle, which determines the average reaction rate, is $\langle T \rangle=\int_0^Tdt \mathcal{P}(n=0,T,\rho_0)$.

Previously, the target survival has been extensively studied, by exploiting single particle results, in the case when the particles are noninteracting RWs. The probability that the target survives until a long time $T$ decays exponentially in time,
\begin{equation}\label{steady}
-\ln{\mathcal P}(n=0,T,\rho_0)\simeq T s(\rho_0),
\end{equation}
with the decay rate \cite{9,10,11,12,13,14,15,16,17}
\begin{equation}
s(\rho_0)= 4\pi CD_0 \rho_0,
\label{soutrw}
\end{equation}
where $C$ is the electrical capacitance of a conductor whose shape is $\Omega$. For a sphere of  radius  $R$ one has $C=R$.
As shown in\cite{MVK},  the long-time expression (\ref{steady}) holds for interacting lattice gases as well, and the steady-state MFT calculations yield model-specific $s(\rho_0)$. Here is a scheme of the calculations. As one can show,  the stationary particle flux, optimal for survival, vanishes everywhere \cite{surv,MVK}. In other words, the
fluctuating contribution to the optimal flux exactly counterbalances the deterministic contribution, thus preventing the particles from being absorbed. One is left with finding the optimal  density profile. Upon the ansatz $\mathbf{J}=0$ and $\rho=\rho(\mathbf{x})$ in Eq.~(\ref{path1}), the action $\mathcal{S}$ becomes proportional to $T$, and the problem reduces to minimizing the \emph{action rate} functional
\begin{equation}\label{lag}
\mathfrak{s}\left[\rho\left(\mathbf{x}\right)\right]=\int d^3\mathbf{x}\frac{\left[D(\rho)\nabla \rho\right]^2}{2\sigma(\rho)},
\end{equation}
subject to the boundary conditions (\ref{blang}) and (\ref{blang3}).
It is convenient to make the transformation  $u(\mathbf{x})=f\left[\rho\left(\mathbf{x}\right)\right]$,
where \cite{conv,MVK}
\begin{equation}
f(\rho)=\int_0^\rho dz \frac{D(z)}{\sqrt{\sigma (z)}}.\label{transform1}
\end{equation}
We denote the inverse function, $f^{-1}$, by $F$.
The transformation (\ref{transform1}) reduces the minimization problem to solving the Laplace's equation
\begin{equation}
\nabla^2 u=0. \label{ustatnd}
\end{equation}
Returning to the original variables, the solution is given in terms of the effective electrostatic potential around a conductor with boundary $\Omega$ kept at unit voltage $\phi\left(\mathbf{x}\right)$. In simple cases (\textit{e.g.}, when $\Omega$ is a disk, a sphere or a spheroid), $\phi(\mathbf{x})$ can be found explicitly \cite{jackson} . The stationary density profile, optimal for the particle survival, is a function of this potential alone:
\begin{equation}\label{rho0}
 \rho(\mathbf{x})=F\{f(\rho_0)[1-\phi(\mathbf{x})]\}.
 \end{equation}
The action rate~(\ref{lag}), evaluated over the solution (\ref{rho0}), yields the decay rate $s(\rho_0)$ entering Eq. (\ref{steady}). It is given by the electrostatic energy created by a conductor $\Omega$ held at voltage $f(\rho_0)$:
\begin{equation}
 s\left(\rho_0\right)= 2\pi  C f^2(\rho_0).
\label{sout}
\end{equation}

Remarkably, the entire effect of interactions is encoded in
the density dependence $f(\rho_0)$, coming from the nonlinear transformation~(\ref{transform1}). The geometry dependence is universal for all gases of this class and is given by the capacitance $C$. When specialized to the RWs, Eq.~(\ref{sout}) reduces to Eq.~(\ref{soutrw}), as to be expected.

\begin{figure}
\includegraphics[width=0.45\textwidth,clip=]{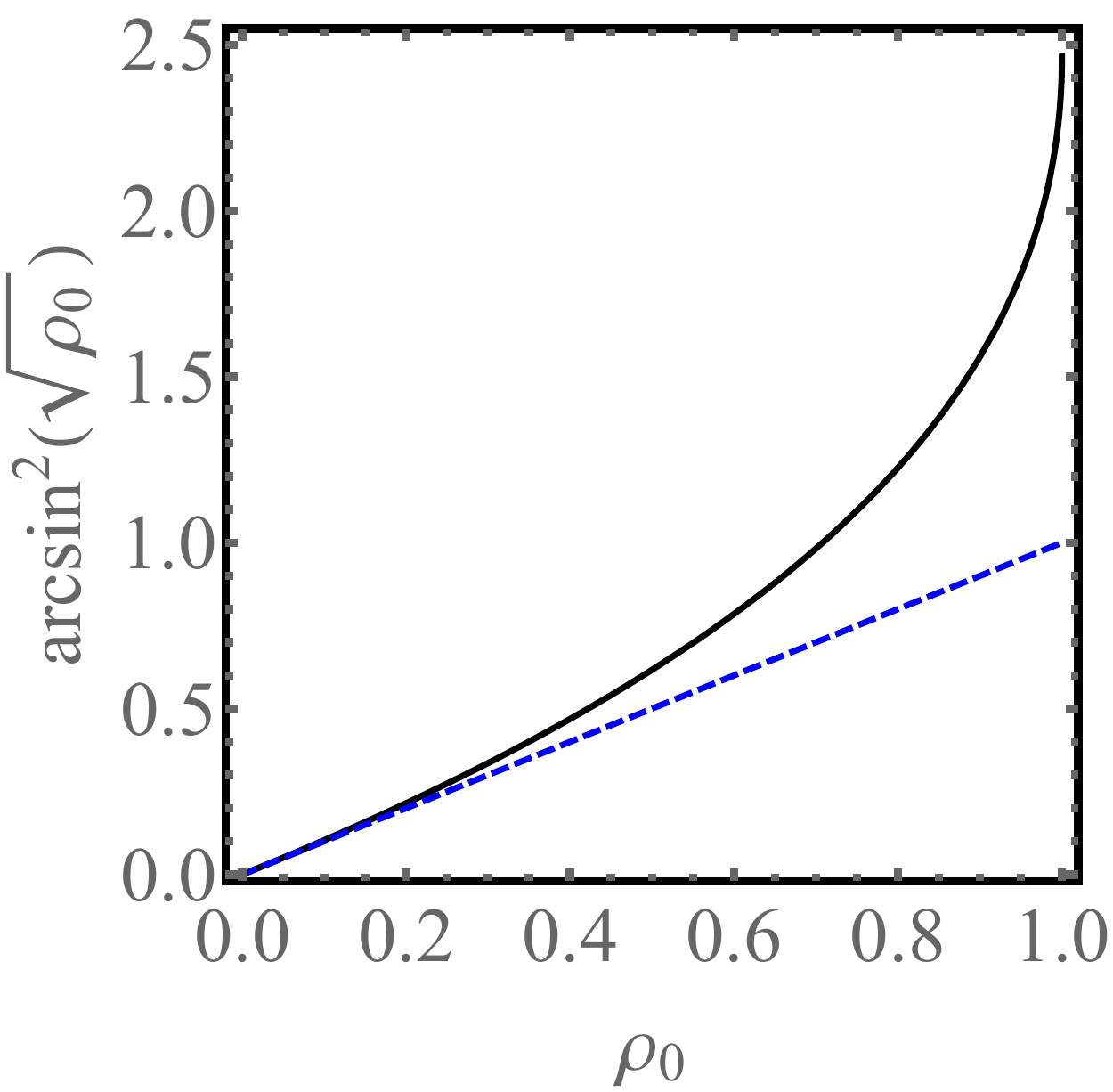}
\includegraphics[width=0.45\textwidth,clip=]{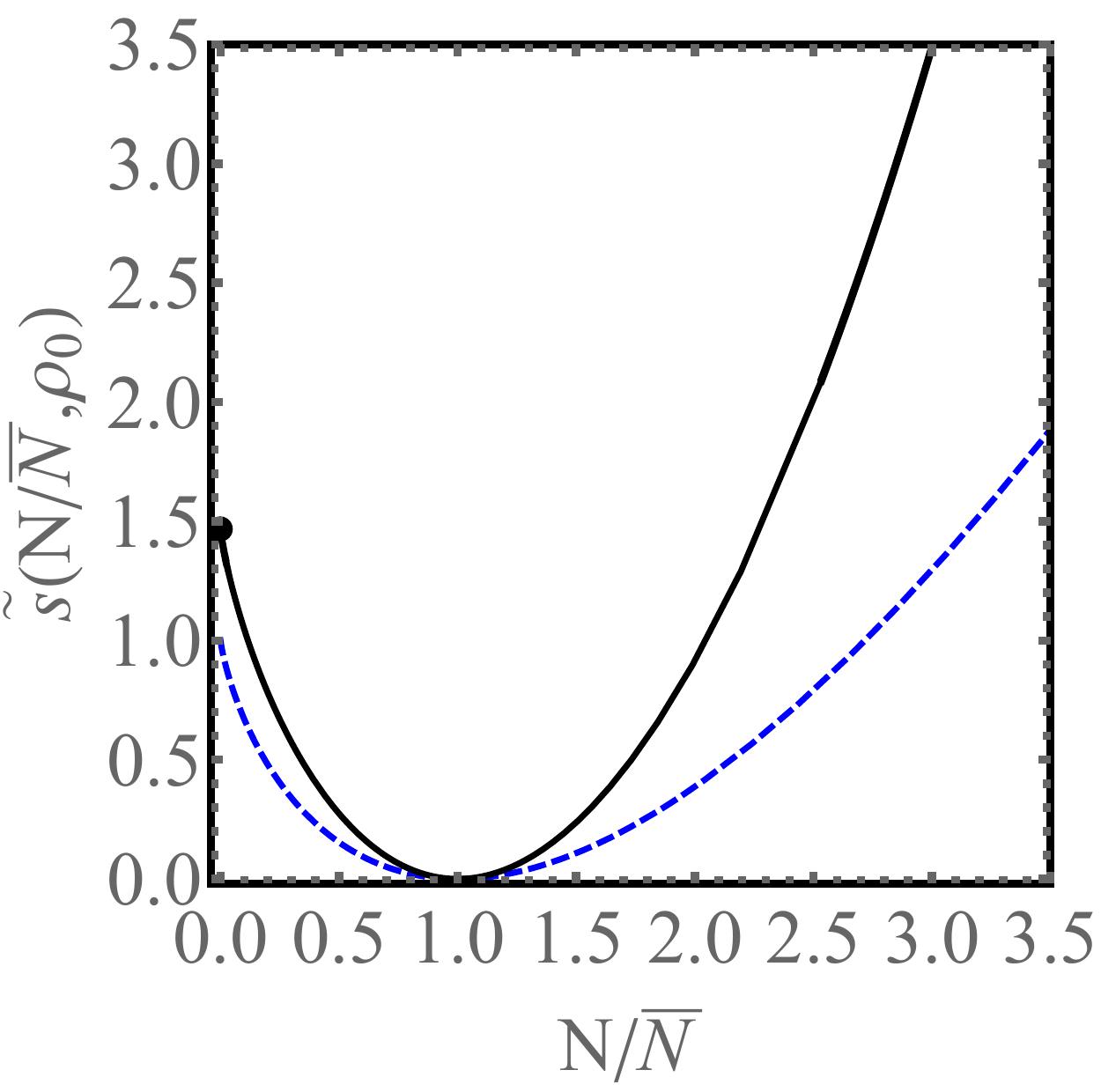}
	\caption{Left panel: the function $\arcsin^2\sqrt{\rho_0}$ which describes the density dependence of the decay rate, Eq.~(\ref{soutssep}), of the target survival probability for the SSEP. The straight line is the same quantity for the RWs, Eq.~(\ref{soutrw}). Right panel: the large-deviation function $\tilde{s}\left(N/\bar{N},\rho_0\right)$ from Eq.~(\ref{fullssep}), which describes the full statistics of absorption in the exterior problem for the SSEP with $\rho_0=0.75$. The dashed line is the same quantity for the RWs, see Eq.~(\ref{pois}). The fat point at $N=0$ shows the survival probability~(\ref{soutssep}).}
	\label{t}
\end{figure}

For the SSEP  Eqs.~(\ref{transform1}) and (\ref{sout}) yield
\begin{equation}
s(\rho_0)=4\pi C D_0 \arcsin^2(\sqrt{\rho_0}).\label{soutssep}
\end{equation}
This decay rate is larger than that of the Rws~(\ref{soutrw}), as to be expected because of the effective
mutual repulsion of the particles, see the left panel of Fig.~\ref{t}. Earlier works \cite{kuz,cum1,cum2,seki1,seki2} on the target survival for the SSEP only established some bounds on $\mathcal{P}(n=0,T,\rho_0)$.

\subsection{Full statistics of absorption}\label{exfull}
When conditioning on arbitrary $n>0$, we impose the constraint $\oint_{\mathbf{x}\in\Omega}\mathbf{J}\cdot\hat n=n$.
The stationary version of  Eq.~(\ref{lang}) is $\nabla \cdot \mathbf{J} = 0$. This fact, alongside with the boundary conditions and the additional assumption that the field $\mathbf{J}$ is irrotational, uniquely defines $\mathbf{J}$  \cite{irot,irot2}. It is given in terms of the average steady-state flux field
\begin{equation}
\bar{\mathbf{J}}\left(\mathbf{x}\right)=-D(\bar{\rho})\nabla\bar{\rho},\label{j1}
\end{equation}
which is also solenoidal and irrotational but obeys  $\oint_{\mathbf{x}\in\Omega}\bar{\mathbf{J}}\cdot\hat n=\bar{n}$. Then it follows that the optimal absorption flux is simply
\begin{equation}
\mathbf{J}=\frac{n}{\bar{n}}\bar{\mathbf{J}},\label{j}
\end{equation}
see also\cite{fullabsorb,shpiel}. The average flux (\ref{j1}) is given by the effective electrostatic potential $\phi\left(\mathbf{x}\right)$, defined in the previous section \cite{berg,multi}:
\begin{equation}
\bar{\mathbf{J}}=V\left(\rho_0\right)\nabla\phi\left(\mathbf{x}\right),\quad V(\rho_0)\equiv\int_{0}^{\rho_0}D(z)dz.\label{j2}
\end{equation}
This potential plays the role of the natural ``spatial coordinate" of the problem.  The optimal density field $\rho$ is a function of the potential alone, and the problem is effectively one-dimensional with respect to this coordinate \cite{shpiel,multi,fullabsorb}:
\begin{equation}
\rho\left(\mathbf{x}\right)=\rho_1\left[\phi\left(\mathbf{x}\right)\right].
\end{equation}
The function $\rho_1\left(u\right)$ is defined on the segment $u\in[0,1]$ and obeys the boundary conditions $\rho_1\left(0\right)=\rho_0$ and $\rho_1\left(1\right)=0$.
It is to be determined by a one-dimensional variational problem which involves the minimization of the action~(\ref{actionmain}). Upon the ansatz $\mathbf{J}=(n/\bar{n})\bar{\mathbf{J}}$ and $\rho\left(\mathbf{x}\right)=\rho_1\left[\phi\left(\mathbf{x}\right)\right]$ in Eq.~(\ref{path1}), the action becomes proportional to $T$, so one needs to minimize an action rate functional. After some algebra the problem is reduced to minimizing a one-dimensional functional in terms of $\rho_1\left(u\right)$\cite{fullabsorb,shpiel,multi}:
\begin{equation}
\mathfrak{s}\left[\rho_1\left(u\right)\right]=4\pi C\times\int_0^1\frac{\left\{\frac{n}{4\pi C}+D\left[\rho_1\left(u\right)\right]\rho_1^{\prime}\left(u\right)\right\}^2}{2\sigma\left[\rho_1\left(u\right)\right]}du\label{s1},
\end{equation}
where, as in the previous section, $C$ is the capacitance of the absorbing domain. The same functional appears in the context of the long-time statistics of the current in a lattice gas on a segment, driven by two reservoirs with different gas densities at the segment's ends \cite{bd}. In the latter setting, the probability of having the current $n$  decays exponentially in time, $-\ln\mathcal{P}_{\text{1d}}\left(n,T,\rho_0\right)\simeq Ts_{\text{1d}}\left(n,\rho_0\right)$. The action rate $s_{\text{1d}}\left(n,\rho_0\right)$ is simply related to the action rate, obtained by minimization of Eq.~(\ref{s1}) $s(n,\rho_0)$:
\begin{equation}
s\left(n,\rho_0\right)=4\pi C s_{\text{1d}}\left(\frac{n}{4\pi C},\rho_0\right) .\label{fulls}
\end{equation}
This sets a universal relation between the different problems \cite{fullabsorb,shpiel,multi}.
The geometry enters only through the capacitance of the domain.

The one-dimensional problem is exactly solvable in quadratures \cite{bd,fullabsorb}. For the SSEP, the result can be written as\cite{fullabsorb}
\begin{equation}
-\ln\mathcal{P}\left(N,T,\rho_0\right)\simeq\bar{N}\tilde{s}\left(\frac{N}{\bar{N}},\rho_0\right),\label{fullssep}
\end{equation}
where $\bar{N} = 4 \pi C D_0 \rho_0 T$. The function $\tilde{s}(N/\bar{N},\rho_0)$ is shown in the right panel of Fig. \ref{t}. In the limit of $\rho_0\to 0$ the function $\tilde{s}(N/\bar{N},\rho_0)$ describes the RWs and corresponds to the $N \gg 1$ limit of the Poisson distribution with mean $\bar{N}$:
\begin{equation}\label{pois}
\tilde{s}\left(\frac{N}{\bar{N}},\rho_0\to 0\right)=\frac{N}{\bar{N}}\ln\frac{N}{\bar{N}}-\frac{N}{\bar{N}}+1.
\end{equation}

\subsection{Multiple absorbing patches}
Ref.\cite{multi} considered particle absorption by \emph{multiple} patches $\Omega_{i},\,i=1,2,\dots,s$, distributed on an otherwise reflecting boundary, see the left panel of Fig.~\ref{hure}.
The results brought some surprises. To start with, the optimal particle flux field, conditioned on a specified joint absorption statistics $\left\{n_i\right\}_{i=1}^s$, exhibits a large-scale \emph{vorticity} $\vec{\omega}=\mathbf{\nabla}\times{\mathbf{J}}\neq 0$ \cite{multi}. The vorticity emerges even when the particles are non-interacting RWs, and for any geometry, as long as there are more than one absorbing patch. This makes the problem more involved as one should consider a joint variational problem for the flux and the density given by Eq.~(\ref{actionmain}). A simplification comes when considering the statistics of typical, small fluctuations, $\delta n_i=n_i-\bar{n}_i \ll \bar{n}_i$, of the absorption currents around their mean values. Here one can linearize the MFT equations around the mean values $\bar{\rho}$ and $\bar{\mathbf{J}}$. The resulting solution\cite{multi} describes a multivariate Gaussian distribution
\begin{equation}\label{mainresult}
{\mathcal P} \simeq
\frac{T^{s/2}}{(2\pi)^{s/2}\,{[\text{det}\,\boldsymbol C]^{1/2}}}\exp\left(-\frac{T}{2}\sum_{i,j=1}^{s}\delta n_iC_{ij}^{-1}\delta n_j\right).
\end{equation}
Here  $\boldsymbol C$ is an $s\times s$ positive-definite symmetric matrix which depends on $\rho_0$ and on the geometry of the problem, but is independent of time. Equation~(\ref{mainresult}) suffices for the evaluation of the variance of the
joint probability distribution \cite{KrMe,multi}. Each diagonal element of $\boldsymbol C$
describes the variance of the current into the corresponding patch:
\begin{equation}\label{vari}
\overline{\delta n_i^2} = \frac{C_{ii}}{T},
\end{equation}
where the overline denotes averaging with respect to the Gaussian distribution (\ref{mainresult}).
The off-diagonal elements of $\boldsymbol C$ describe cross-correlations between the currents into different patches:
\begin{equation}\label{corr}
\overline{ \delta n_i\delta n_j } = \frac{C_{ij}}{T}.
\end{equation}
The optimal density field can again be presented via an electrostatic analogue which involves $s$ characteristic potentials $\phi_i\left(\mathbf{x}\right)$. Each of the potentials appears when the corresponding conducting patch $\Omega _i$ is held at unit voltage, the rest of the conducting patches are grounded, and the Neumann boundary condition is specified at the reflecting part of the boundary. The potentials $\phi_i$-s can be found explicitly in simple cases \cite{jackson,multi}. The covariance matrix $\boldsymbol C$ is given in terms of the volume integrals involving the characteristic potentials:
\begin{equation}
C_{ij}= \int d\mathbf{x} \,\sigma (\bar{\rho})\nabla\phi_i\cdot\nabla\phi_j.\label{mat}
\end{equation}
Remarkably, general properties of the cross-correlations turn out to be independent of the system's geometry, and are determined solely by the functions $D(\rho)$ and $\sigma(\rho)$ \cite{multi}. Of course, there are no cross-correlations if the particles do not interact.
What is the sign of cross-correlations for an interacting gas? As Ref.\cite{multi} showed, if
\begin{equation}
D(\bar{\rho})\sigma^{\prime\prime}(\bar{\rho})<D^{\prime}(\bar{\rho})\sigma^{\prime}(\bar{\rho}), \label{condition}
\end{equation}
for any value of $\bar{\rho}\in[0,\rho_0]$,
then the currents into different patches $i\neq j$ are all anti-correlated, $\overline{ \delta n_i\delta n_j }<0$, regardless of the system's geometry. In particular, this is always true for the SSEP.

Interestingly, the same condition (\ref{condition}) guarantees the validity of the additivity principle (that is, stationarity of the optimal density profile in the long-time limit) for an arbitrary value of current \cite{ber,main}, and also determines the sign of the two-point \emph{density} correlation function \cite{bertinistat,tridib}, in single-current systems.

\section{The interior problem}

For many non-interacting particles the theory is based on the well-established single-particle results\cite{keller1,bress,beni,bookz,russians,RoKim,many2}
As in the exterior problem, the long-time survival probability in this case
decays exponentially in time, $-\ln\mathcal P\left(T,\rho_0\right)\simeq Ts\left(\rho_0\right)$. The geometry dependence of $s\left(\rho_0\right)$ is, however,  different\cite{keller1,bress,beni,bookz,russians,RoKim,many2}:
\begin{equation}\label{suas}
s(\rho_0)=D_0 \mu_0^2 \rho_0V,
\end{equation}
where  $V$ is the domain's volume, and $\mu_0^2$ is the principal eigenvalue of the eigenvalue problem $\nabla^2 \Psi+\mu^2 \Psi=0$ inside the domain with the mixed boundary conditions $\Psi(\mathbf{x} \in  \Omega_a,t) = \nabla\Psi(\mathbf{x} \in \Omega_r,t)\cdot\hat{n}=0$.

What happens for interacting particles? As in section \ref{exsurv}, the optimal flux field, conditioned on the survival of all particles, vanishes identically. As a result, we can determine the optimal gas profile for survival by minimizing the same action rate functional~(\ref{lag}), but now the integration is carried over the space inside the domain. One distinct feature of the interior survival problem is conservation of the total number of particles, which enters the variational problem as a constraint,
\begin{equation}\label{cons}
\int d^3\mathbf{x}\,\rho(\mathbf{x})=\rho_0 V
\end{equation}
and calls for a Lagrange multiplier $\Lambda$.  The transformation of variables (\ref{transform1}) proves useful in the interior case as well. The resulting Euler-Lagrange equation for $u$ has the form of
a non-linear Poisson equation \cite{surv},
\begin{equation}
\nabla^2 u +  \Lambda \frac{d F(u)}{du}=0, \label{ustatnd2}
\end{equation}
with the mixed boundary conditions \cite{boundary},
\begin{equation}
u(\mathbf{x}\in{\Omega_a})=0 ,\quad\nabla u(\mathbf{x}\in{\Omega_r})\cdot\hat{n}=0.
\label{bqu2}
\end{equation}
For the RWs Eq.~(\ref{transform1}) yields
$F\left( u\right)=u^2/2D_0$, and Eq.~(\ref{ustatnd2}) becomes the Helmholtz equation
\begin{equation}\label{helm2}
\nabla^2 u+\mu^2 \,u=0 ,
\end{equation}
with $ \mu^2\equiv\Lambda/D_0$ playing the role of the eigenvalue. The minimum action is achieved for the fundamental mode,
and the resulting expression for $\mathfrak{s}[u(\mathbf{x})]= s(\rho_0)$ reproduces the result quoted in Eq.~(\ref{suas})\cite{surv,narrow}.
For the SSEP, upon rescaling $U= \sqrt{2/D_0}\, u$ and $C= \Lambda/D_0$, Eq.~(\ref{ustatnd2}) becomes the stationary sine-Gordon equation
\begin{equation}
	\nabla^2 U +  C\sin\,U=0. \label{ustatssepnd1}
\end{equation}

\subsection{Particle survival inside a fully absorbing domain}

\begin{figure}
	\includegraphics[width=0.50\textwidth,clip=]{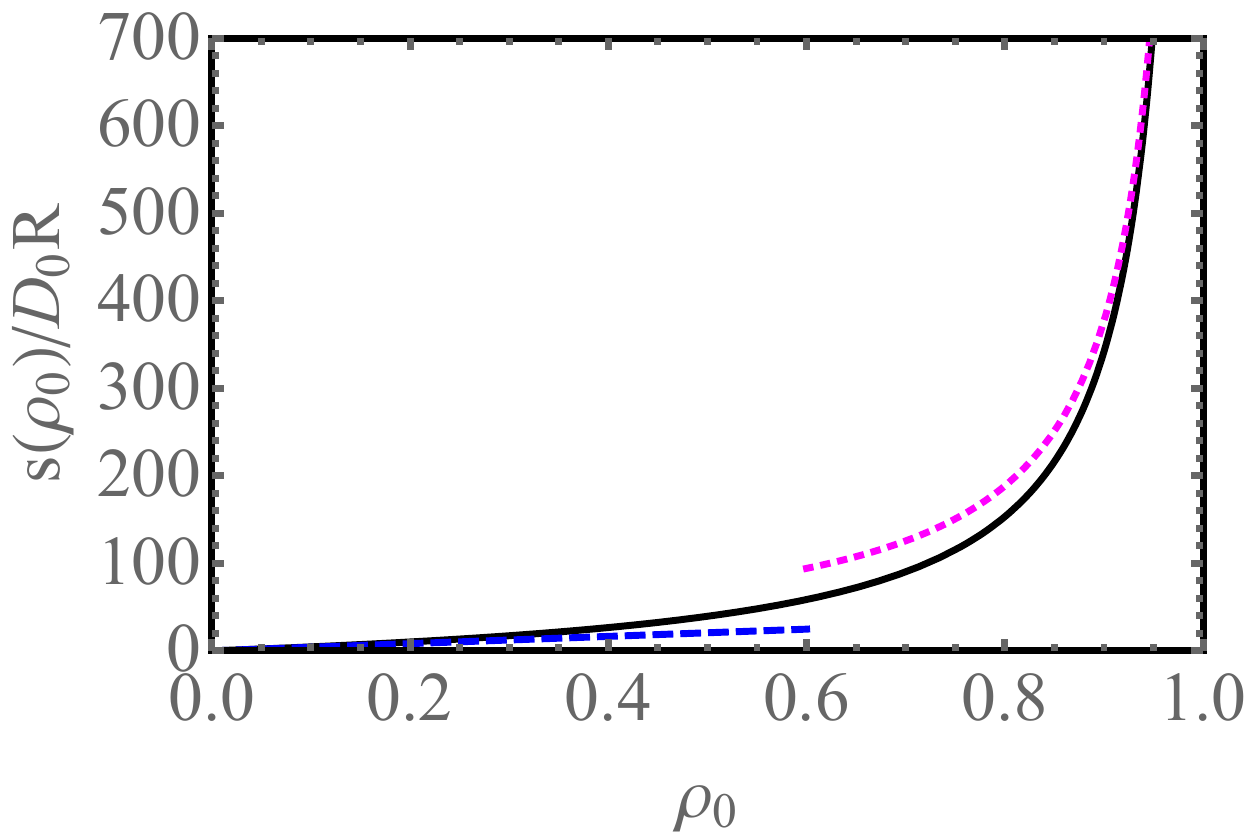}
	\caption{Solid line: the rescaled action rate $s(\rho_0)/D_0R$ for a sphere, vs. $\rho_0$, obtained by numerically solving Eq.~(\ref{ustatssepnd1}) and using Eq.~(\ref{lag}). Also shown are the low-density asymptotic~(\ref{ssphere}) (dashed line) and the high-density asymptotics~(\ref{divergence3d}) (dotted line).}
	\label{s3d}
\end{figure}

Equation~(\ref{ustatssepnd1}) can be solved exactly in some simple geometries. Among them are a one-dimensional segment (where the problem is exactly solvable for any gas model) and a rectangle \cite{surv}. For a sphere of radius $R$ one can solve Eq.~(\ref{ustatssepnd1}) numerically, and also explore analytically the low- and high-density limits. In the dilute limit $\rho_0 \ll 1$ one reproduces the RWs result (\ref{suas}) which becomes
\begin{equation}\label{ssphere}
s_{\text{RWs}}(\rho_0)=\frac{4\pi^3}{3}RD_0 \rho_0,
\end{equation}
At the other extreme, $\rho_0 \to 1$, the stationary optimal density profile $\rho$ stays very close to $1$ across most of the domain, and drops to $0$ in a narrow boundary layer of characteristic width $\delta =1-\rho_0$ along the domain boundary. As a result, the problem becomes effectively one-dimensional in the direction normal to the domain boundary.
The solution for this one-dimensional problem can be found exactly, and the action rate, Eq.~(\ref{lag}), mostly comes from the boundary layer. The final result, for a general domain shape, is
\begin{equation}\label{highs}
s(\rho_0) \simeq \frac{D_0 A^2 }{V\left(1-\rho_0\right)},
\end{equation}
where $A$ is the surface area of the boundary.
For a sphere of radius $R$ one obtains
\begin{equation}
s(\rho_0)\simeq
\frac{12\pi D_0R}{1-\rho_0}. \label{divergence3d}
\end{equation}
Figure~\ref{s3d} shows the numerically found $s(\rho_0)/D_0R$,  alongside with the asymptotics (\ref{ssphere})  and (\ref{divergence3d}) \cite{surv}.

\subsection{Narrow escape of interacting particles}
In the narrow escape problem, particles can escape only through a small escape hole $\Omega_a$, of size $\epsilon \ll L$, see the right panel of Fig.~\ref{hure}.
The mean escape time of the first particle (MET) in this setting determines the rates of important processes in molecular and cellular biology \cite{diffapfere,rna,den,bress}.

For the non-interacting RWs, one can evaluate $\mu_0^2$ in Eq.~(\ref{suas}) perturbatively with respect to the small parameter $\epsilon/L$.
In the leading  order $\mu_0^2$ can be expressed through the electrical capacitance $C_{\epsilon}$ of the conducting patch $\Omega_a$ in an otherwise empty space: $\mu_0^2\simeq 2\pi C_{\epsilon}/V$ \cite{rayleigh}. The capacitance  $C_{\epsilon}$ scales as $\epsilon$. When $\Omega_a$ is a disk  of radius $\epsilon$, one has $C_{\epsilon}=2\epsilon/\pi$ \cite{jackson}.
The resulting survival probability decay rate (\ref{suas}) is \cite{RoKim,narrow}
\begin{equation}\label{asuas}
s(n_0,\epsilon)\simeq 2\pi C_{\epsilon}  D_0\rho_0.
\end{equation}
For interacting particles one can exploit the small
parameter $\epsilon/L$ in a similar way \cite{narrow}. The leading-order contribution to the action rate (\ref{lag}) comes from only a vicinity of the escape hole. That is, to leading order in $\epsilon/L$, the solution for a finite domain coincides with the one for
a gas of particles occupying the infinite half-space on one side of an infinite reflecting plane with the hole $\Omega_a$ on it. This reduces the problem to the \emph{unconstrained} minimization procedure of the exterior survival problem of section \ref{exsurv}.
The solution is therefore given by Eq.~(\ref{rho0}) where $\phi(\mathbf{x})$ is the electrostatic potential  of a conducting patch $\Omega_a$ kept at unit voltage on an otherwise insulating infinite plane. If the escape hole is a circle, $\phi(\mathbf{x})$ can be found explicitly \cite{jackson}. Then
Eq.~(\ref{lag}) yields the decay rate of the non-escape probability to order $\epsilon/L$:
\begin{equation}
s\left(\rho_0,\epsilon\right) \simeq \pi  C_{\epsilon} f^2(\rho_0).
\label{sout2}
\end{equation}
As in the exterior survival problem, the gas-specific interactions are encoded in
the density dependence $f(\rho_0)$, whereas the geometry dependence $C_{\epsilon}$ is universal. To leading order it only depends on the shape of the escape hole and is independent of the domain shape.
For the SSEP  inside a domain with a small circular escape hole
of radius $\epsilon$ one obtains\cite{narrow}
\begin{equation}
s(\rho_0,\epsilon)\simeq 4D_0\epsilon \arcsin^2(\sqrt{\rho_0}).\label{soutssep2}
\end{equation}
The density dependence of Eq.~(\ref{soutssep2}) is the same as in Eq.~(\ref{soutssep}), see Fig.~\ref{t}.
As argued in \cite{narrow} , for the SSEP with random initial condition, the exponential decay of ${\mathcal P}$ with time $T$ holds as soon as $T$ is much longer than the diffusion time across the escape hole \cite{russians,RoKim,narrow}. For sufficiently low gas densities, $\rho_0\epsilon^3 \ll 1$, the MET of the first particle, $\langle T \rangle$ is also much longer than this diffusion time, and is thus given by $\langle T \rangle\simeq1/s(\rho_0,\epsilon)$ \cite{RoKim,narrow}.

\section{Short-time statistics: non-stationary fluctuations}\label{nonstat}
At short times, $T \ll L^2/D\left(\rho_0\right)$, the particle absorption statistics, in both exterior and interior settings, strongly depend on the initial condition, whereas the optimal density profile explicitly depends on time \cite{MVK}.
Here we must return to the full time-dependent MFT formulation given by Eqs.~(\ref{path1}) and (\ref{actionmain}). A universal simplification comes from the fact that, for very short times, the domain size is irrelevant. As a result, the process is effectively one-dimensional in the direction normal to the absorbing part of the boundary \cite{narrow,MVK,RoKim}, and
the absorption statistics can be expressed through that of a gas on the infinite half-line $x>0$, with absorbing boundary conditions at $x=0$ \cite{1d1,1d2,1d3,1d4,1d5,1d6,1d7,1d8,1d9,MVK}.
The particle \emph{survival} probability on the half-line is well studied \cite{1d1,1d2,1d3,1d4,1d5,1d6,1d7,1d8,1d9,MVK}. For the RWs with random initial conditions one obtains $-\ln {\mathcal P_{1d}}^{\text{(rand)}} \simeq (2/\sqrt{\pi})  \rho_0 \sqrt{D_0T}$. The corresponding result for the deterministic (or quenched) setting differs by a numerical factor \cite{1d2,MVK} .
Remarkably, for the SSEP one obtains the same stretched-exponential
decay with time as for the RWs: $-\ln {\mathcal P_{1d}} \simeq s_{\text{1d}}(\rho_0)\sqrt{D_0T}$, but the density dependence $s_{\text{1d}}(\rho_0)$ is now different for the different initial conditions.
The low-density expansion of $s_{\text{1d}}(\rho_0)$  was recently calculated: $s^{\text{(rand)}}_{\text{1d}}(\rho_0) = (2/\sqrt{\pi})[\rho_0 + (\sqrt{2}-1) \rho_0^2 + \dots]$ \cite{MVK,Santos}. For larger $\rho_0$, $s_{\text{1d}}(\rho_0)$ can be computed numerically \cite{MVK}. To evaluate the survival probability ${\mathcal P}(T,\rho_0)$, one should multiply the action $s_{\text{1d}}(\rho_0)$ by
the surface area $A$ of the absorbing part $\Omega_a$ \cite{MVK,narrow,RoKim} :
\begin{equation}
-\ln {\mathcal P}(T,\rho_0)\simeq A s_{\text{1d}}(\rho_0) \sqrt{D(\rho_0)T} .
\label{shorttimeint}
\end{equation}
At sufficiently high densities, $\rho_0L^3\gg 1$, the short-time expression (\ref{shorttimeint}) suffices for the evaluation of the MET. Indeed, in this regime the MET is much shorter then the diffusion time across the domain, $L^2/D_0$, and we obtain $\langle T \rangle \simeq 2 [A^2 D(\rho_0) s_{\text{1d}}^2(\rho_0)]^{-1}$\cite{RoKim,narrow}. For the narrow escape problem the relevant diffusion time scale is $\epsilon^2/D_0$ \cite{RoKim,narrow}.
As an example, consider a circular absorbing patch of radius $\epsilon$. In this case we have for the RWs $\langle T_{\text{RWs}} \rangle^{\text{(rand)}}\simeq (2\pi  D_0 \rho_0^2 \epsilon^4)^{-1}$ \cite{RoKim,narrow}. For the SSEP the MET is shorter because of the effective particle repulsion: $\langle T \rangle^{\text{(rand)}}\simeq \langle T_{\text{RWs}} \rangle^{\text{(rand)}}\left[1-2 (\sqrt{2}-1) \rho_0 + \dots\right]$.

\section{Summary}

The fluctuating hydrodynamics and macroscopic fluctuation theory provide a simple, general and versatile framework for the study of kinetics of diffusion-controlled reactions in multi-particle systems where intra-reactant interactions are important. We demonstrated the versatility of these approaches in several exterior and interior settings of particle survival, absorption and escape. More complicated settings and geometries can be also considered. The approach can be extended in different directions. For example, it can accommodate simple reactions among, and a finite lifetime of, the particles \cite{ElgartKamenev,reac1,reac2,hurtado2,reac3}.

\section{Acknowledgments}

This research was supported
by the United States-Israel Binational Science
Foundation (BSF) (Grant No. 2012145) and  by the Israel Science Foundation (Grant No. 807/16).



\begin{thebibliography} {99}
	
	\bibitem{MVK}
	B. Meerson, A. Vilenkin, and P. L. Krapivsky, Phys. Rev. E \textbf{90}, 022120 (2014).
	
\bibitem{surv}

T. Agranov, B. Meerson, and A. Vilenkin,
Phys. Rev. E \textbf{93}, 012136 (2016).

\bibitem{fullabsorb} B. Meerson, J. Stat. Mech. P05004 (2015).
	
\bibitem{multi}
T. Agranov and B. Meerson, Phys. Rev. E \textbf{95}, 062124 (2017).

\bibitem{narrow}
T. Agranov and B. Meerson,
Phys. Rev. Lett. \textbf{120}, 120601 (2018).

\bibitem{bress}
P. C. Bressloff and J. M. Newby, Rev. Mod. Phys. \textbf{85}, 135 (2013).

\bibitem{keller1}
M. J. Ward and J. B. Keller, SIAM J. Appl. Math. \textbf{53}, 770 (1993).

\bibitem{russians}
I. V.  Grigoriev, Y. A. Makhnovskii, A. M. Berezhkovskii, and V. Y. Zitserman, J. Chem. Phys. \textbf{116}, 9574 (2002).


\bibitem{beni}
O. B\'{e}nichou and R. Voituriez, Phys. Rep. \textbf{539}, 225 (2014).

\bibitem{bookz}
D. Holcman and Z. Schuss, \textit{Stochastic Narrow Escape in Molecular and Cellular Biology}
(Springer, New York, 2015).

\bibitem{rev}
T. Chou and M. R. D'Orsogna, in ``\textit{First-Passage Phenomena and Their Applications}",
edited by R. Metzler,  G. Oshanin, and S. Redner (World Scientific, Singapore 2013).

\bibitem{rice}
S. A. Rice, \textit{Comprehensive Chemical Kinetics} (Elsevier, Amsterdam, 1985).

\bibitem{osh} C. Mejıa-Monasterio, G. Oshanin, and G. Schehr, J.Stat.Mech. (2011) P06022.

\bibitem{red}
B. Meerson and S. Redner, Phys. Rev. Lett. \textbf{114}, 198101 (2015).

\bibitem{berg}
	H. C. Berg and E. M. Purcell, Biophys. J. \textbf{20}, 193 (1977).
	
\bibitem{bergbook}
	H. C. Berg, \textit{Random Walks in Biology} (Princeton University Press, Princeton, USA, 1993).

\bibitem{LL} L. D. Landau and E. M. Lifshitz, \textit{Statistical Physics} (Pergamon Press, London, 1958).

\bibitem{Spohn}
	H. Spohn, \textit{Large-Scale Dynamics of Interacting Particles} (Springer-Verlag, New York, 1991).
	
\bibitem{MFTreview}
	
	L. Bertini, A. De Sole, D. Gabrielli, G. Jona Lasinio, C. Landim. Rev. Mod. Phys. \textbf{87}, 593 (2015).

\bibitem{KL}
C. Kipnis and C. Landim, \textit{Scaling Limits of Interacting Particle Systems} (Springer, New York, 1999).

\bibitem{Liggett}
T. M. Liggett, \textit{Stochastic Interacting Systems: Contact, Voter,
	and Exclusion Processes} (Springer, New York, 1999).

\bibitem{Paulbook} P. L. Krapivsky, S. Redner, and E. Ben-Naim, \textit{A Kinetic View of Statistical Physics} (Cambridge University Press, Cambridge, 2010).

\bibitem{MSR} P. C. Martin, E. D. Siggia, and H. A. Rose, Phys. Rev. A \textbf{8}, 423 (1973).


\bibitem{layers}  The full time-dependent solution of the problem develops two narrow boundary layers in time, at
$t=0$ and $t=T$. They only give a subleading contribution to the action (\ref{actionmain}), see \textit{e.g.} \cite{surv,MVK}.

\bibitem{bd}

T. Bodineau and B. Derrida, Phys. Rev. Lett. \textbf{92}, 180601 (2004).


\bibitem{9}
G. Zumofen, J. Klafter, and A. Blumen, J.Chem.Phys. \textbf{79}, 5131 (1983).

\bibitem{10} M. Tachiya, Radiat. Phys. Chem. \textbf{21}, 167 (1983).

\bibitem{11}
S. Redner and K. Kang, J. Phys. A \textbf{17}, L451 (1984).

\bibitem{12} A. Blumen, G. Zumofen, and J. Klafter, Phys. Rev. B \textbf{30}, 5379(R) (1984).

\bibitem{13} A. Blumen,J. Klafter, and G. Zumofen, in \textit{Optical Spectroscopy of Glasses}, edited by I. Zchokke (Reidel, Dordrecht, 1986), p. 199.

\bibitem{14} S. F. Burlatsky and A. A. Ovchinnikov, Sov. Phys. JETP \textbf{65}, 908 (1987).

\bibitem{15} G. Oshanin, O. Benichou, M. Coppey, and M. Moreau, Phys. Rev. E \textbf{66} 060101(R) (2002).

\bibitem{16} R. A. Blythe and A. J. Bray, Phys. Rev. E \textbf{67}, 041101 (2003).

\bibitem{17} A. J. Bray, S. N. Majumdar, and G. Schehr, Adv. Phys. \textbf{62}, 225 (2013).

\bibitem{conv}
Convergence of the integral (\ref{transform1})
puts some limitations on the behavior of $D(\rho)$ and $\sigma(\rho)$ at small densities. As an example, let $D(\rho\rightarrow0)\sim\rho^{\alpha}$ and $\sigma(\rho\rightarrow0)\sim\rho^{\beta}$. Then the integral converges at $\rho\rightarrow0$ if and only if $2\alpha-\beta+2>0$. This condition holds in the examples we consider here.

\bibitem{jackson}
J. D. Jackson,  \textit{Classical Electrodynamics} (Wiley, New York, 1999).

\bibitem{kuz}
V. Kuzovkov and E. Kotomin, Phys. Rev. Lett. \textbf{72}, 2105 (1994).

\bibitem{cum1}
S. F. Burlatsky, M. Moreau, G. Oshanin, and A. Blumen,
Phys. Rev. Lett. \textbf{75}, 585 (1995).

\bibitem{cum2}
D. P. Bhatia, M. A. Prasad, and D. Arora
Phys. Rev. Lett. \textbf{75}, 586 (1995).

\bibitem{seki1}
K. Seki and M. Tachiya, Phys. Rev. E. \textbf{80}, 041120 (2009).

\bibitem{seki2}
K. Seki, M. Wojcik, and M. Tachiya,
J. Chem. Phys. \textbf{134}, 094506 (2011).

\bibitem{irot}
One also needs to use the fact that the component of the flux, transverse to the domains boundary, vanishes. This property  holds for the optimal irrotational flux field.
\bibitem{irot2}
Within linearized MFT equations,  the vorticity of the flux field vanishes \cite{multi} .
It would be very interestig to find out whether a nonzero vorticity can emerge, in nontrivial geometries, beyond small fluctuations.

\bibitem{shpiel}
E. Akkermans, T. Bodineau, B. Derrida and O. Shpielberg, EPL \textbf{103}, 20001 (2013).

\bibitem{KrMe} P. L. Krapivsky and B. Meerson,  Phys. Rev. E \textbf{86}, 031106 (2012).

\bibitem{ber}

L. Bertini, A. De Sole, D. Gabrielli, Jona-Lasinio and C. Landim, J. Stat. Phys. \textbf{123}, 237 (2006).

\bibitem{main}
O. Shpielberg and E. Akkermans, Phys. Rev. Lett. \textbf{116}, 240603 (2016).


\bibitem{bertinistat}

L. Bertini, A. De Sole, D. Gabrielli, G. Jona-Lasinio, and C. Landim  J. Stat. Phys. \textbf{135}, 857 (2009).

\bibitem{tridib}

T. Sadhu and B. Derrida,  J. Stat. Mech. (2016) 113202.


\bibitem{RoKim} S. Ro and Y. W. Kim, Phys. Rev. E \textbf{96}, 012143 (2017).

\bibitem{many2}
K. Basnayake, C. Guerrier, Z. Schuss, and D. Holcman, arXiv:1711.01330.

\bibitem{boundary}
The condition $u(\mathbf{x}\in{\Omega_a})=0$ is inherited from
$\rho(\mathbf{x}\in{\Omega_a})=0$ due to the definition (\ref{transform1}). The condition $\nabla u(\mathbf{x}\in{\Omega_r})\cdot\hat{n}=0$ results from a boundary term that appears when minimizing the action (\ref{lag}).

\bibitem{diffapfere}
D. Coombs, R. Straube, and M. Ward, SIAM. J. Appl. Math. \textbf{70}, 302 (2009).

\bibitem{rna}
S. A. Gorski, M. Dundr, and T. Misteli, Curr. Opin. Cell Biol. \textbf{18}, 284 (2006).


\bibitem{den}
D. Holcman, Z. Schuss, and E. Korkotian, Bio. J. \textbf{87}, 81 (2004).

\bibitem{rayleigh}
J. W. S. Baron Rayleigh \textit{The Theory of Sound}, 2nd ed. (Dover, New York, 1945), vol. 2.

\bibitem{1d1}
M. Tachiya, Radiat. Phys. Chem. \textbf{21}, \textbf{167} (1983).

\bibitem{1d2} G. Zumofen, J. Klafter, and A. Blumen, J. Chem. Phys. \textbf{79}, 5131 (1983).

\bibitem{1d3}
S. Redner and K. Kang, J. Phys. A Math. Gen. \textbf{17},  L451 (1984).


\bibitem{1d4}
A. Blumen, G. Zumofen, and J. Klafter, Phys. Rev. B \textbf{30}, 5379(R) (1984).

\bibitem{1d5}
S. F. Burlatsky and A. A. Ovchinnikov, Sov. Phys. JETP \textbf{65}, 908 (1987).

\bibitem{1d6}
R. A. Blythe and A. J. Bray, Phys. Rev. E \textbf{67}, 041101  (2003).

\bibitem{1d7}
J. Franke and S. N. Majumdar, J. Stat. Mech. P05024 (2012).

\bibitem{1d8}
A. J. Bray, S. N. Majumdar, and G. Schehr, Adv. Phys. \textbf{62}, 225 (2013).

\bibitem{1d9}
S. Redner and B. Meerson, J. Stat. Mech. P06019 (2014).

\bibitem{Santos} J. E. Santos and G. M. Sch\"{u}tz, Phys. Rev. E \textbf{64}, 036107 (2001).

\bibitem{ElgartKamenev}
V. Elgart and A. Kamenev, Phys. Rev. E \textbf{70}, 041106 (2004).

\bibitem{reac1}
T. Bodineau and M. Lagouge, J. Stat. Phys. \textbf{139}, 201 (2010).

\bibitem{reac2} B. Meerson and P. V. Sasorov, Phys. Rev. E \textbf{83}, 011129 (2011).

\bibitem{hurtado2} P. I. Hurtado, A. Lasanta, and A. Prados, Phys. Rev. E \textbf{88}, 022110 (2013).

\bibitem{reac3} B. Meerson, J. Stat. Mech. P05004 (2015).
	

\end{thebibliography}
\end{document}